\documentclass[useAMS,usenatbib]{mn2e}
\usepackage{graphicx,graphics,amsmath}
\usepackage{latexsym}

\begin{document}

\title{Possibility of conversion of neutron star to quark star in presence 
of high magnetic field}

\author[Ritam Mallick and Monika Sinha]
{Ritam Mallick$^{x}$ and Monika Sinha $^{y}$ \\ 
$^{x}$ Department of Physics, Indian Institute of Science,
Bangalore 560012, INDIA \\
ritam@physics.iisc.ernet.in \\
$^{y}$ Institut f\"ur Theoretische
Physik, J. W. Goethe-Universit\"at, D-60438 Frankfurt am
Main, Germany \\ 
Department of Physics, Indian Institute of Science,
Bangalore 560012, INDIA \\ 
sinha@th.physik.uni-frankfurt.de}

\maketitle

\begin{abstract}
Recent results and data suggests that high magnetic field in neutron 
stars (NS) strongly affects the characteristic (radius, mass) of the
star. They are even separated as a class known as magnetars, for whom 
the surface magnetic field are greater than $10^{14}$ G. In this work 
we discuss the effect of such high magnetic field on the phase transition
of NS to quark star (QS). We study the effect of magnetic 
field on the transition from NS to QS including the magnetic 
field effect in equation of state (EoS). The inclusion of 
the magnetic field increases the range of 
baryon number density, for which the flow velocities of the matter in 
the respective phase are finite. The magnetic 
field helps in initiation of the conversion process. The velocity of the 
conversion front however decreases due to the presence of
magnetic field, as the presence of magnetic field reduces 
the effective pressure (P). The magnetic field of the star gets decreased by 
the conversion process, and the resultant QS has lower magnetic 
field than that of the initial NS.
\end{abstract}

stars: neutron, equation of state, gravitation, hydrodynamics, stars: magnetic fields, shock waves

\section{Introduction} \label{intro}

The discovery of radio pulsar \citep{hewish}, 
brought about the theoretical proposition of NS
\citep{bz} to much attention. Inside a NS the matter, composed 
of neutrons, protons, electrons, and sometimes
muons, is in a highly dense state, whose density may be as high as 
$3-10$ times normal nuclear matter (NM) saturation density ($n_0$). 
Naturally, the constituent particles
therein interact via strong forces, forming a non-ideal nuclear 
fluid. However, the nature of the strong interaction at this 
high density is not well understood yet. There are many theoretical
models of nuclear 
matter at high density describing different EoSs for the matter.
Based on different EoSs, different
mass-radius relations of NS are obtained, which can only be justified by 
comparing the theoretical results with observed properties of NS. 

The possibility of emergence and existence of QS, containing deconfined
quarks in their free states has been intensively 
discussed in the literature \citep{al,olinto}.
If the QS consists only of almost equal number of free up (u), down (d)
and strange (s) quarks it is termed as strange star (SS) 
\citep{key-3,al}, otherwise
a more general star consisiting of all kinds of quarks in their free 
states is called a QS.
In this work our final EoS is for that of a SS.
The measured periods and spin down rates of soft-gamma repeaters 
(SGR) and of anomalous X-ray pulsars (AXP), and the observed 
X-ray luminosities of AXP, indicate that some NSs have 
extremely high surface magnetic fields, as large as $10^{14}-10^{15}$ G  
\citep{key-4,key-5} which are known as magnetars. Furthermore, if these 
sources are the central engine of 
gamma-ray bursts (GRBs), as suggested in \citep{key-6,key-7}, their 
surface magnetic field 
might even be larger. The discovery of 
magnetars has triggered a growing interest in the study of the structure, 
dynamics and the evolution of NS with large magnetic 
fields, which has raised a number of interesting issues.

It is known that magnetic field plays important role in the astrophysical 
phenomena, such as supernovae, GRBs, galaxy jet, and so on. 
Recently, there is a growing consensus in explaining SGRs via the magnetar 
model \citep{key-4}. Magnetars are believed to be NS with strong 
magnetic field which is responsible for the observed flare activity. Three 
giant flares, SGR 0526-66, SGR 1900+14 and SGR 1806-20, have been detected so 
far. This huge amount of energy can be explained by the presence of a strong 
magnetic field whose strength is estimated to be larger than 
$4\times10^{14}$ G. In the late part of the flares, a careful analysis revealed 
the existence of characteristics quasi-periodic oscillations (QPO) 
\citep{key-8}. It is not clear whether they are associated to crustal modes, 
or to modes of the magnetic field or both; if the spacing between the observed 
frequencies would be explained, one may gain information on the internal 
structure of the star \citep{key-9}.

The existence of a magnetar motivates to study the effects of 
strong magnetic field on NS properties. A strong magnetic field
affects, the structure of a NS through its influence on
the underlying metric \citep{bbgn,cpl} and EoS
through the Landau quantization of charged particles and then 
the interaction of magnetic moments of charged particles with 
the magnetic field. For the 
NM with a $n$-$p$-$e$ system, the effect of 
magnetic field was studied by several authors \citep{cbs,yz,bpl1,czl,wmkksg}. 
However, as discussed earlier the composition of the 
core of a NS is very uncertain, and different EoSs have been proposed
to describe the matter at such extreme condition. The matter in the star 
may contain only deconfined quarks, which are known as SS,
or the hyperons may 
appear, making hyperonic matter. 
The effect of magnetic field 
on quark matter using the MIT bag model has been studied  
earlier \citep{chakra,gc,fmro}. There are other
models of quark matter with phenomenological density dependent
quark masses \citep{fowler,chak,d98,lxl}. Broderick et al. \citep{bpl2} 
studied the effect of strong magnetic field on hyperonic matter, where 
the field strength does not depend on density. 
However, in reality, we expect the field 
strength should be higher at core than at surface of a NS. Therefore, 
the field strength should vary with radius, and hence with density.

So the study of magnetic field in NS may provide with new different 
results and will help in understanding the basic properties of NS in a much 
better way. The magnetic field will also play an important role in the 
conversion of NS to SS. The NS may convert to a SS by several different ways.
A few possible mechanisms for the production
of SQM in a NS have been discussed by Alcock  et al 
\citep{al}.
The conversion from hadron matter to quark matter is expected to start
as the star comes in contact with a seed of external strange quark
nugget. Another mechanism for the initiation of the conversion process was
given by Glendenning \citep{key-10}. It was suggested there that a
sudden spin down of the star may increase the density at its core
thereby triggering the conversion process spontaneously.

Conversion of NM to SQM has been studied by
several authors, which are discussed in detail by 
Bhattacharyya et al \citep{abhi} 
and for brevity we do not discuss them here.

In this paper we plan to study the effect of the density dependent magnetic
field EoS on the conversion front.
We will write down the Rankine-Hugoniot condition for the matter 
velocities and solve 
them with the EoS derived in presence of magnetic field. In 
presence of magnetic field, both pressure ($P$) and energy density 
($\varepsilon$)
of the system are affected. This will indirectly give the 
effect of magnetic field on the conversion front. The 
paper is arranged in the following way: In the next section we 
construct our model for the magnetic field dependent EoS.
In section \ref{rankine} we will discuss the kinematics of the phase transition
for the magnetic field dependent EoS. In section \ref{dynamic} we will 
discuss about 
the propagation of the front along the star and in the final section we 
will summarize our results. 

\section{Model}\label{model}

First we construct the magnetic field induced EoS.
We use nonlinear Walecka 
model \citep{wal}, which has been successful in describing the nuclear 
ground state properties and elastic scattering \citep{walecka,chin,serot,sw}. 
In addition to the model, here we consider the possibility
of appearance of hyperons ($\Lambda, \Sigma^-, \Sigma^0, \Sigma^+,
\Xi^-, \Xi^0$) and muons ($\mu^-$) at higher density. 

The detail of the calculation is similar to that of Sinha et al \citep{sinha}
and for brevity we only mention the important results of the model here.
For the magnetic field inclusion we choose the gauge to be, 
$A^{\mu}\equiv(0,-y{\cal B},0,0)$, ${\cal B}$ being 
the magnitude of magnetic field
and $eQ$ the charge of the particle with $e$ the positive 
unit of charge. For this 
particular gauge choice ${\vec {\cal B}}={\cal B}\hat{z}$. 
In the presence of magnetic field, the motion of the charged particles
is Landau quantized in the perpendicular direction to the magnetic
field. The momentum in the $x$-$y$ plane is quantized and hence the energy
in the $n$th Landau level is given by
\begin{equation}
E_n =  \sqrt{p_z^2 + m^2 + 2 n e |Q| {\cal B}}.
\end{equation}

With the above consideration we can write down the 
total energy density of matter as
\begin{eqnarray}
\varepsilon &=&  \frac12 m_\sigma^2 \sigma^2 + U(\sigma)
 + \frac12 m_\omega^2 \omega^{0^2}~+~\frac12 m_\rho^2 \rho_3^{0^2}
\nonumber \\
&& + \sum_N 
\frac1{8 \pi^2} \left(2 {p_F^{(N)}\mu_N^{*^3}}
 - p_F^{(N)}m_N^{*^2}\mu_N^* 
%\right. \nonumber \\ &&  \left. 
- m_N^{*^4}~\ln \left[\frac{p_F^{(N)}+\mu_N^*}{m^*_N} \right]
\right) \nonumber\\
&& +  \frac{e|Q|{\cal B}}{(2\pi)^2} \sum_C 
\sum_{n=0}^{n_{max}} (2-\delta_{n,0})
\nonumber \\ && \times
\left(p_C(n)\mu_C^* 
%\right. \nonumber \\ && \left.
+~( m_C^{*^2} + 2ne|Q|{\cal B}) 
\ln \left[\frac{p_C(n)+\mu_C^*}{\sqrt {(m_C^{*2} + 2ne|Q|{\cal B})}}\right]\right) \nonumber\\
&& + \frac{e|Q|{\cal B}}{(2\pi)^2} \sum_{l=e,\mu} \sum_{n=0}^{n_{max}}
(2-\delta_{n,0}) 
\nonumber \\ && \times
\left(p_l(n)\mu_l 
%\right. \nonumber \\ && \left. 
+~(m_l^2 + 2ne|Q|{\cal B}) 
\ln \left[\frac{p_l(n)+\mu_l}{\sqrt{(m_l^2 + 2ne|Q|{\cal B})}}\right]\right) 
\nonumber \\
&& + \frac {{\cal B}^2}{8\pi},
\label{enden}
\end{eqnarray}
where $N$ denotes charge neutral baryons, $C$ charged baryons 
and $l$ leptons. 
$\psi_B,\psi_l,\sigma,\omega$ and $\rho$ are fields of baryons, leptons, 
$\sigma$-mesons, 
$\omega$-mesons and $\rho$-mesons, with masses $m_B, m_l, m_\sigma,
m_\omega$ and $m_\rho$ respectively, $g_{\sigma B},
g_{\omega B}$ and $g_{\rho B}$ are coupling constants for interactions
of $\sigma, \omega$ and $\rho$ mesons respectively with the baryon
$B$. $U(\sigma)$ is the scalar self interaction term \citep{glend8,bb}.
We define $p(n)~=~\sqrt{p_F^2 - 2ne|Q|{\cal B}}$, where $p_F$ 
is the Fermi momentum.

The total $P$ is then 
\begin{equation}
P=\sum_B \mu_B n_B + \sum_l \mu_l n_l - \varepsilon.
\label{press}
\end{equation} 

For SQM we employ one of the realistic EoS with density dependent
quark masses given by Dey et al. \citep{d98}. In this model 
quarks interact among themselves via Richardson potential \citep{rich}.
The matter is composed of $u$, $d$ and $s$ quarks and electrons.
In the presence of magnetic field the energy density is then 
given by
\begin{equation}
\varepsilon = \varepsilon_k + \varepsilon_p + \varepsilon_e
+ \frac {{\cal B}^2}{8\pi},
\end{equation}

where 
\begin{eqnarray}
\varepsilon_k =  \frac{3e{\cal B}}{(2\pi)^2} \sum_i |Q|_i
\sum_{n=0}^{n_{max}} (2-\delta_{n,0})
\nonumber \\  \times
\left(p_i(n)\mu_i^* 
%\right. \nonumber \\ && \left.
+~( m_i^2 + 2ne|Q|_i{\cal B}) 
\ln \left[\frac{p_i(n)+\mu_i^*}{\sqrt {(m_i^2 + 2ne|Q|_i{\cal B})}}\right]\right) \nonumber\\
\end{eqnarray}
is the kinetic energy density of quarks with $\mu_i^*~=~\sqrt
{m_i^2+p_F^{(i)^2}}$, $i=u,d,s$,

\begin{eqnarray}
\varepsilon_p = -\frac1{(2\pi)^4}\sum_{i,j}(e{\cal B})^2 |Q|_i|Q|_j
\int_0^{p_i(n_i)}\int_0^{p_j(n_j)}\int_0^{2\pi} V(q)
\nonumber \\ 
\times F(p_i,p_j,m_i,m_j,\phi)\, dp_z^{(i)}\, dp_z^{(j)}\, d\phi
\end{eqnarray}
is the potential energy density with 

\begin{eqnarray}
F(p_i,p_j,m_i,m_j) = \frac{(E_i+m_i)(E_j+m_j)}{4E_iE_j} \times \nonumber \\
\left\{1 + \frac{p_i^2p_j^2}{(E_i+m_i)^2(E_j+m_j)^2}
+\frac {2 {\mathbf p_i} \cdot {\mathbf p_j}}{(E_i+m_i)(E_j+m_j)} \right\}
\end{eqnarray}
and 
\begin{eqnarray}
\varepsilon_e = 
\frac{e|Q|_e{\cal B}}{(2\pi)^2} \sum_{n=0}^{n_{max}}
(2-\delta_{n,0})
\nonumber \\ \times
\left(p_e(n)\mu_e
%\right. \nonumber \\ && \left.
+~(m_e^2 + 2ne|Q|_e{\cal B})                                               
\ln \left[\frac{p_e(n)+\mu_l}{\sqrt{(m_e^2 + 2ne|Q|_e{\cal B})}}\right]\right) 
\end{eqnarray}                                                             
is the energy density of electrons.                                        
The total pressure is then                                                 
                                                                           
\begin{equation}                                                           
P=\sum_i \mu_i n_i +  \mu_e n_e - \varepsilon.                             
\label{press1}                                                              
\end{equation}                                                             
                                                                           
The relation between energy density and pressure describes the             
EoS of the matter.       

As the configuration of the 
magnetic field inside a star is not known, following previous work 
\citep{bcs} we adopt a magnetic field profile
\begin{equation}
{\cal B}\left(n_b/n_0\right)={\cal B}_S+{\cal B}_C\left\{1-e^{-\beta \left(
\frac {n_b}{n_0} \right)^\gamma}\right\},
\label{magprfl}
\end{equation}
where $\beta$ and $\gamma$ are two parameters determining the 
magnetic field profile with given ${\cal B}_S$ and ${\cal B}_C$,
and $n_b$ is the total baryon number density.
We assume that the magnetic field at the center is higher than at the 
surface by a few order of magnitude. As magnetars are observed to have
${\cal B}_S$ as large as $10^{15}$ G, in our model we keep ${\cal B}_S$
fixed at $10^{15}$ G and vary the ${\cal B}_C$.
It is found that the effect of magnetic 
field is important for ${\cal B}_C \ge 10^{17}$ G. However, 
for ${\cal B}_C < 10^{18}$ G, the effect is significant only 
when the field reaches ${\cal B}_C$ at a very low density, 
away from the center, and remains almost constant up to center. 
Therefore, we restrict our study with ${\cal B}_C \sim\,10^{18}$ G. 

At lower densities, the matter is composed of only neutrons, 
protons and electrons. Hence, at the low density regime, the 
particles
which are affected by the magnetic field are electrons and protons. 
Since the electrons are highly relativistic, electron Fermi momentum 
is very large compared to electron mass. Therefore, the number 
of occupied Landau levels by electrons is very large, 
even though the field strength under consideration is larger 
than the critical field strength of electron by several orders. 
On the other hand, the field strength under consideration is
very less than the critical field strength of protons. 
Consequently, the number of occupied Landau levels by protons
is also large.  As density increases, the heavier particles 
appear gradually. In addition, the magnetic field increases 
with the increase of density. As a result, the number of occupied
Landau levels gradually decreases for every species. The threshold 
densities for muons and other hyperons to appear for ${\cal B}
_C=10^{18}$ G  are given in Table \ref{threshold}. 
The threshold densities for various species to appear do not differ 
from their respective values when the magnetic field is 
absent.  
\begin{table}[t]
\begin{center}
\begin{tabular}[t]{|c|c|}
\hline
\multicolumn{2}{|c|}{Threshold densities} \\
\hline
$\mu^-$ & 0.9 \\
\hline
$\Lambda$ & 2.6 \\
\hline
$\Xi^-$ & 3.1 \\
\hline
$\Xi^0$ & 5.7 \\
\hline
\end{tabular}
\caption{Threshold densities for muons and hyperons to appear in units
of $n_0$.}
\label{threshold}
\end{center}
\end{table}

\begin{figure}[t]
\begin{center}
\includegraphics[width=2.5 in]{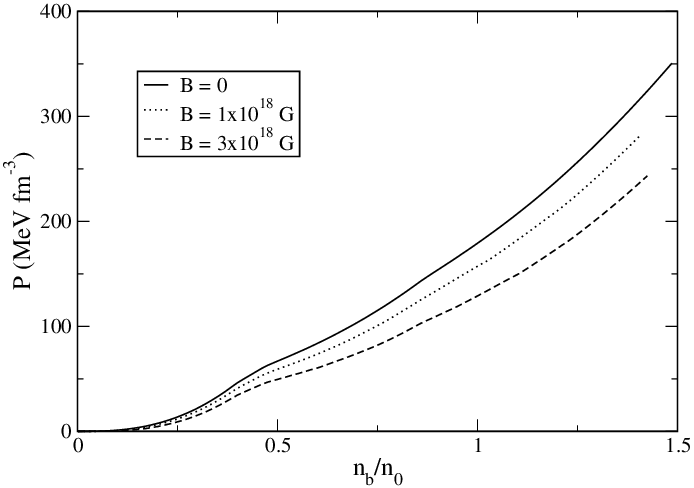}
\caption{Variation of $P$ as a function of normalized baryon number 
density. 
The solid curve is for without magnetic field. The 
dotted and dashed curves correspond to ${\cal B}_C=1\times 10^{18}$ G, 
and $3\times 10^{18}$ G respectively.}
\label{eos}
\end{center}
\end{figure}

In fig. \ref{eos} we show the EoSs with and without magnetic 
field when the magnetic field profile corresponds to $\beta
=0.1$ and $\gamma=1$ [see Eq. (\ref{magprfl})]. We observe 
that the EoS becomes softer 
with the increase of ${\cal B}_C$. Here 
we should mention that, when the field strength is high enough, 
the field energy and the field pressure are not negligible. 
In calculating the EoS, we thus add this contribution too which 
is necessary to construct the structure of NS. 

\section{Rankine-Hugoniot condition} \label{rankine}

We heuristically assume the existence of a combustive phase transition 
front. Using the macroscopic
conservation conditions, we examine the range of densities for which
such a combustion front exists. We next study the outward propagation
of this front through the model star by using the hydrodynamic (\textit{i.e.}
Euler) equation of motion and the equation of continuity for the energy
density flux \citep{key-17}. 
Let us now consider the physical situation where a combustion front
has been generated in the core of the NS. This front propagates
outwards through the NS with a certain hydrodynamic velocity. 
In the following, we denote all the physical
quantities in the hadronic sector by subscript $1$ and those in the
quark sector by subscript $2$.

Quantities on opposite sides of a front
are related through the energy density, the momentum density and the
baryon number density flux conservation. In the rest frame of the
combustion front, these conservation conditions can be written as
\citep{key-17,key-23,key-23a}:

\begin{equation}
\omega_{1}v_{1}^{2}\gamma_{1}^{2}+P_{1}=\omega_{2}v_{2}^{2}\gamma_{2}^{2}+P_{2},
\label{2}\end{equation}

\begin{equation}
\omega_{1}v_{1}\gamma_{1}^{2}=\omega_{2}v_{2}\gamma_{2}^{2},
\label{3}\end{equation}
and
\begin{equation}
n_{1}v_{1}\gamma_{1}=n_{2}v_{2}\gamma_{2}.
\label{4}\end{equation}

In the above three Rankine-Hugoniot conditions 
$v_{i}$ $(i=1,2)$ is the velocity, 
$\gamma_{i}=\frac{1}{\sqrt{1-v_{i}^{2}}}$ is the
Lorentz factor, $\omega_{i}=\varepsilon_{i}+P_{i}$ is the specific enthalpy
of the respective phases.

The velocities of the matter in the two phases, given by eqns. 
(\ref{2}-\ref{4}), are written as \citep{key-23}:

\begin{equation}
v_{1}^{2}=\frac{(P_{2}-P_{1})(\varepsilon_{2}+P_{1})}{(\varepsilon_{2}
-\varepsilon_{1})(\varepsilon_{1}+P_{2})},
\label{5}\end{equation}

and \begin{equation}
v_{2}^{2}=\frac{(P_{2}-P_{1})(\varepsilon_{1}+P_{2})}{(\varepsilon_{2}
-\varepsilon_{1})(\varepsilon_{2}+P_{1})}.
\label{6}\end{equation}

It is possible to classify the various conversion mechanism 
by comparing
the velocities of the respective phases with the corresponding 
sound speed, denoted by $c_{si}$, in these phases. 
For the conversion to be physically possible, velocities should satisfy
an additional condition, namely, $0\leq v_{i}^{2}\leq 1$. 

\begin{figure}
\vskip 0.3in
   \centering
\includegraphics[width=2.5 in]{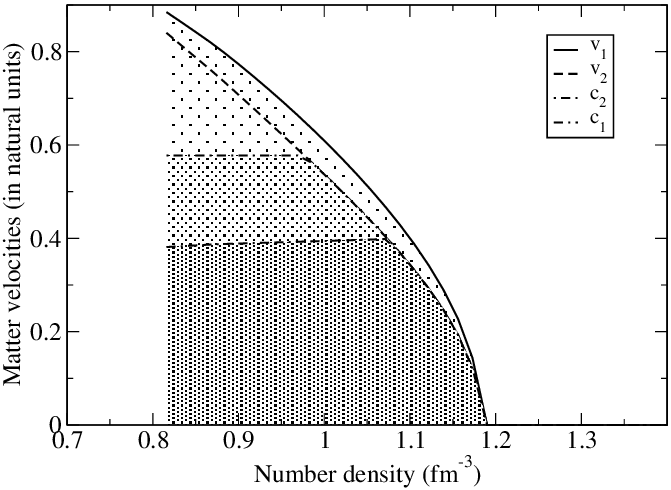}
\caption{Variation of different flow velocities with baryon number density. 
The most densly shaded region correspond to deflagration, moderate shaded 
region correspond to detonation and the most lightly shaded region
correspond to supersonic conversion processes.}
\label{all}
\end{figure}

For the above constructed EoS with magnetic field we first 
plot (fig. \ref{all}) the variation of different matter 
flow velocities as a function of baryon number density. 
We shade the different portion of the graph which represents different 
modes of conversion mechanism. For this EoS, beyond a certain density
(on the lower side) the curve does not comes down. This is because,
below that density the energy and velocity criterion are not satisfied. 
If we now construct a star with this hyperon matter EoS then the minimum 
central density has to be that high where the curve end in the lower 
baryon density region. At that density there is a high chance of 
detonation wave formation as the difference in $v_1$ and $v_2$ is high. 
Whereas at very high 
density a star may be formed but the conversion process would not start. 
This is because at that very high density the star becomes so dense that 
whatever may be the perturbation, it fails to grow to give rise to a wave 
which would start the conversion process. 

\begin{figure}
\vskip 0.3in
   \centering
\includegraphics[width=2.5 in]{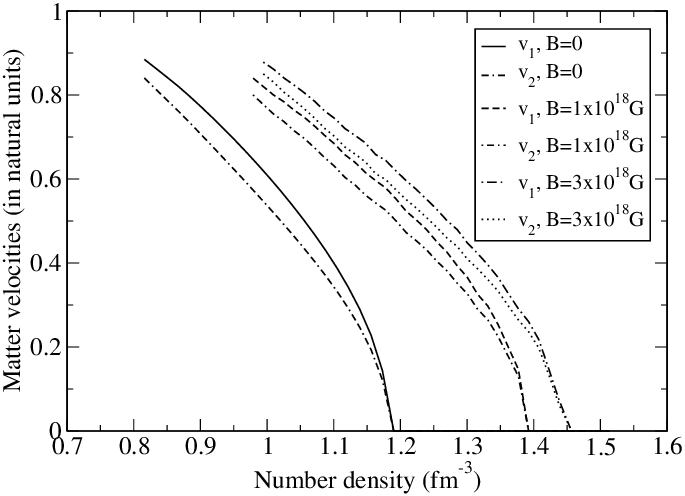}
\caption{Variation of $v_1$ and $v_2$ with baryon number density.
The curves are plotted for the EoS with magnetic field profile corresponding to 
$\beta=0.1$ and $\gamma=1$. The variation is plotted for three cases, one 
without magnetic field and the other two with ${\cal B}_C 
=1\times10^{18}$ G and $3\times10^{18}$ G.}
\label{kin1}
\end{figure}

\begin{figure}
\vskip 0.3in
   \centering
\includegraphics[width=2.5 in]{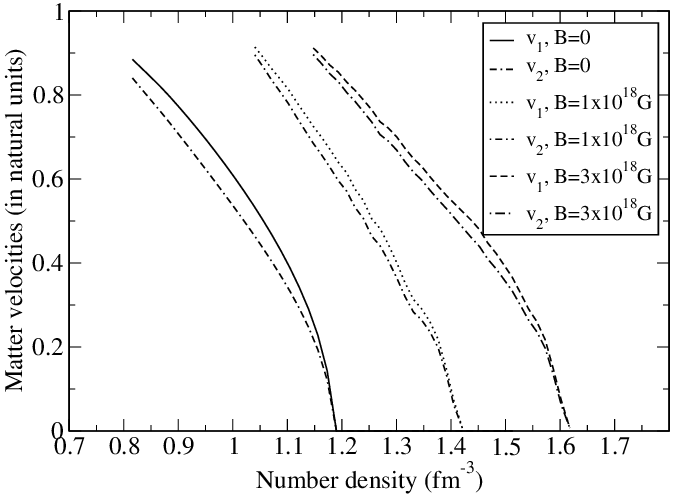}
\caption{Variation of $v_1$ and $v_2$ with baryon number density.
The curves are plotted for the EoS with magnetic field profile corresponding to 
$\beta=0.2$ and $\gamma=2$. The variation is plotted for three cases, one 
without magnetic field and the other two with ${\cal B}_C 
=1\times10^{18}$ G and $3\times10^{18}$ G.}
\label{kin2}
\end{figure}

\begin{figure}
\vskip 0.3in
   \centering
\includegraphics[width=2.5 in]{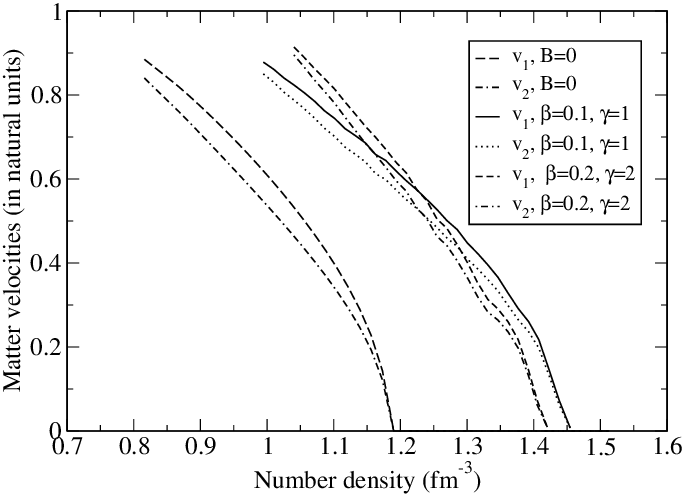}
\caption{Variation of $v_1$ and $v_2$ with baryon number density.
The curves are plotted for the EoSs with magnetic field profiles 
corresponding to 
$\beta=0.1$, $\gamma=1$ and $\beta=0.2$, $\gamma=2$ with ${\cal B}_C =
3\times10^{18}$ G along with the non magnetic case.}
\label{allgb}
\end{figure}

In fig. \ref{kin1} we plot  $v_1$ and $v_2$ for different
${\cal B}_C$ 
corresponding to $\beta=0.1$ and $\gamma=1$.
For comparison we also plot for the non magnetic case. 
We find that the nature of the curve remains same, that is $v_1$ is always 
greater than $v_2$, which means the shock front propagates outward of the 
star. Due to introduction of the 
magnetic field the range of baryon
density, for which the flow velocities are physical, increases. The magnetic 
field decreases the effective $P$ for the same
baryon number density, rendering the matter to be more compressible. That is
for the same density now the matter is less rigid or more compressible now. 
For much higher value of the baryon density where the matter was very dense
previously, there was little chance of shock formation, but now due to the
introduction of the magnetic field there is a finite chance of shock formation.
Therefore the range of baryon density gets much wider. In fig. \ref{kin2} we 
plot the same, but for the case of the EoS with magnetic field 
corresponding to  $\beta=0.2$ and $\gamma=2$. The nature of the graph remains 
the same. 
From eqn. \ref{magprfl}, we can see that $\beta$ and $\gamma$ describes how 
the magnetic field varies from the core to the surface for 
fixed ${\cal B}_C$ and 
${\cal B}_S$ values. It determines the slope of the magnetic field, that is higher 
the $\beta, \gamma$ value larger the slope. Therefore for the same value of 
the magnetic field the range of density, for which the flow velocities are 
finite increases with the increase of $\beta, \gamma$. This becomes much more
clear from fig. \ref{allgb}, where we have plotted for two sets 
of $\beta,\gamma$ value 
for same $\cal B_C$. The slope of $v_1$ and $v_2$ changes with the 
change in slope of the magnetic field configuration. 

\section{Propagation of the front} \label{dynamic}

Next we come to the dynamic picture of the front propagation, that is the 
evolution of the hydrodynamic combustion front along the radius of the star.
To examine such an evolution, we move to a reference
frame in which the NM is at rest. The speed of the combustion
front in such a frame is given by ${v}={-v}_{1}$. The velocity of matter 
in the combustion frame is in natural units, with $c=\hbar=k_{B}=1$.
All the other physical quantities discussed below are also in natural 
units, that is they are converted to $eV$.

Using special relativistic formalism to
study the evolution of combustion front we derive the relevant Eulers
and continuity equation, given by \citep{key-17}:

\begin{equation}
\frac{1}{\omega}(\frac{\partial\varepsilon}{\partial\tau}
+v\frac{\partial\varepsilon}{\partial r})+
\frac{1}{W^{2}}(\frac{\partial v}{\partial r}
+v\frac{\partial v}{\partial\tau})+2\frac{v}{r}=0
\label{7}\end{equation}

and 
\begin{equation}
\frac{1}{\omega}(\frac{\partial P}{\partial r}+
v\frac{\partial P}{\partial\tau})+
\frac{1}{W^{2}}(\frac{\partial v}{\partial\tau}+
v\frac{\partial v}{\partial r})=0,
\label{8}\end{equation}
where, $v=\frac{\partial r}{\partial\tau}$ is the front velocity
in the NM rest frame and $k=\frac{\partial P}{\partial\varepsilon}$
is taken as the square of the effective sound speed in the medium and
$W=1/\gamma_{i}$ is the inverse of Lorentz factor.

The above equations are solved to give a single equation
\begin{equation}
\frac{dv}{dr}=\frac{2vkW^{2}(1+v^{2})}{r[4v^{2}-k(1+v^{2})^{2}]}.
\label{11}\end{equation}

The eqn. (\ref{11}) is integrated, with respect to $r(t)$, starting
from the center towards the surface of the star. Using the above EoS we 
construct a star following the standard
Tolman-Oppenheimer-Volkoff equations \citep{key-17a}. 
The velocity
at the center of the star should be zero from symmetry considerations.
On the other hand, the $1/r$ dependence of the $ \frac {dv}{dr} $,
in eqn. (\ref{11}) suggests a steep rise in velocity near the center of the star. 

We first construct the density
profile of the star for a fixed central density
(here it is $6$ times normal nuclear saturation density). Equations (\ref{5})
and (\ref{6}) then specify the respective flow velocities $v_{1}$
and $v_{2}$ of the nuclear and quark matter in the rest frame of
the front, at a radius infinitesimally close to the center of the
star. This would give the initial velocity of the front ($-v_{1}$),
at that radius, in the NM rest frame. We next start with
eqn. (\ref{11}) from a point infinitesimally close to the center
of the star and integrate it outwards along the radius of the star.
The solution gives the variation of the velocity with the distance
from the center of the star. 
For a static star, being spherically symmetric, the problem is rather simple; 
for a rotating star, however, the asymmetry has to be taken care of. 

As due to rotation the star is no more spherical, but is oblate spheroid,
therefore to describe the star 
we introduce a new parameter $\chi=\cos\theta$, where $\theta$ is the angle 
made with the vertical axis (axis of rotation) of the star. The detail 
general relativistic (GR) calculations can be obtained from Bhattacharyya 
et al. \citep{abhi1}. We here mention only the important results needed 
for our purpose. 
We start with the metric \citep{cst}
\begin{eqnarray}
ds^2 = -e^{\gamma+\rho}dt^2 + e^{2\alpha}(dr^2+r^2d\theta^2) + 
e^{\gamma-\rho}r^2 sin^2\theta(d\phi-\omega dt)^2
\end{eqnarray}
describing the structure of the star, with the four gravitational potentials 
$\alpha, \gamma, \rho$ and $\omega$, which are functions of $\theta$ and $r$ 
only. The Einstein's equations for the potentials are solved through the 
{\bf `rns'} code, with the input of our EoS and a fixed central 
density. Similar hydrodynamic equation for the front propagation can be
written down and solved to get the final equation for the velocity of 
the front. The equation is quite similar to that of
(\ref{11}), only there is an extra term due to GR effect \citep{abhi1}.

\begin{figure}[t]
\vskip 0.3in
   \centering
\includegraphics[width=2.5 in]{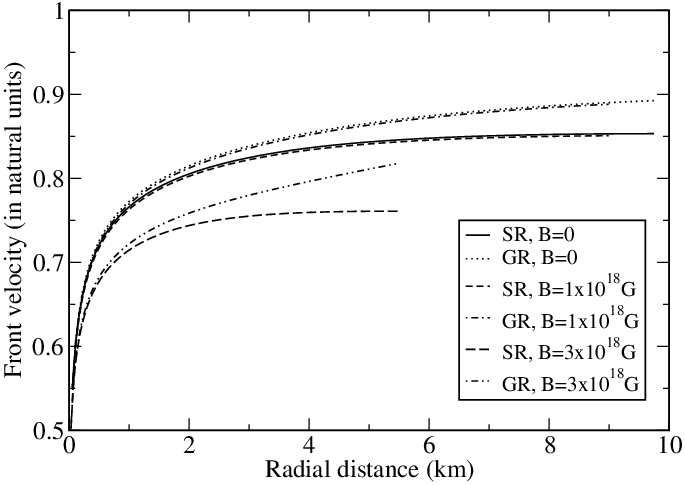}
\caption{Variation of the velocity of the conversion front with radius of the
star. The curves are plotted for the EoS with magnetic field profile 
corresponding to $\beta=0.1$ and $\gamma=1$. The variation is plotted for 
three cases, one 
without magnetic field and the other two with ${\cal B}_C =1\times10^{18}$ G 
and $3\times10^{18}$ G.}
\label{dyn1}
\end{figure}

\begin{figure}
\vskip 0.3in
   \centering
\includegraphics[width=2.5 in]{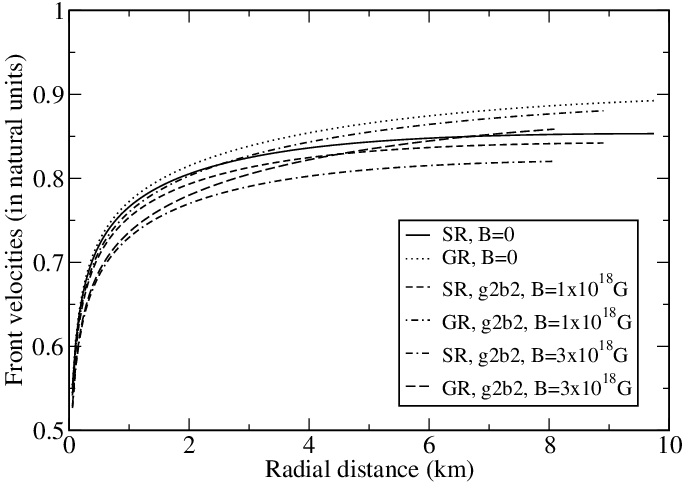}
\caption{Variation of the velocity of the conversion front with radius of the
star. The curves are plotted for the EoS with magnetic field profile 
corresponding to $\beta=0.2$ and $\gamma=2$. The variation is plotted for 
three cases, one 
without magnetic field and the other two with ${\cal B}_C=1\times10^{18}$ G 
and $3\times10^{18}$ G.}
\label{dyn2}
\end{figure}

In fig. \ref{dyn1} we plot the velocity of the conversion 
front along the 
radius of the star in the presence of magnetic field for different
${\cal B}_C$, with field profile corresponding to $\beta=0.1$ 
and $\gamma=1$. We show both results obtained from
special and general relativistic calculations. The figure
shows that the velocity of the front, for all cases,
shoots up near the center and then saturates at a certain velocity
for higher radius. Such a behaviour of velocity near the central point
is apparent from the eqn. (\ref{11}) above. The velocity 
of the front further increases  with the inclusion of GR effect.
The rise in the front velocity is due to the fact that 
now GR effect of the curvature of the front adds up with the $P$ 
contribution which provides the thrust to the propagation front. We can 
further observe that as the magnetic field increases the velocity of the 
front decreases, this is due to the fact that the magnetic contribution 
(${\cal B}^2/8 \pi$) acts in the opposite direction of the matter $P$, reducing 
the effective $P$ term. This $P$ is the fact which provides the 
thrust to the propagation front, and thus the velocity of the 
front decreases 
with the decrease of effective $P$. Fig. \ref{dyn2} shows the same 
nature only now we plot for the EoS with magnetic field 
profile corresponding to $\beta=0.2$ and $\gamma=2$. The difference in the 
graphs is only due to the difference in the nature of the magnetic field 
(mainly the slope of the growth of the field which is governed by 
$\beta$ and $\gamma$).

\begin{figure}
\vskip 0.3in
   \centering
\includegraphics[width=2.5 in]{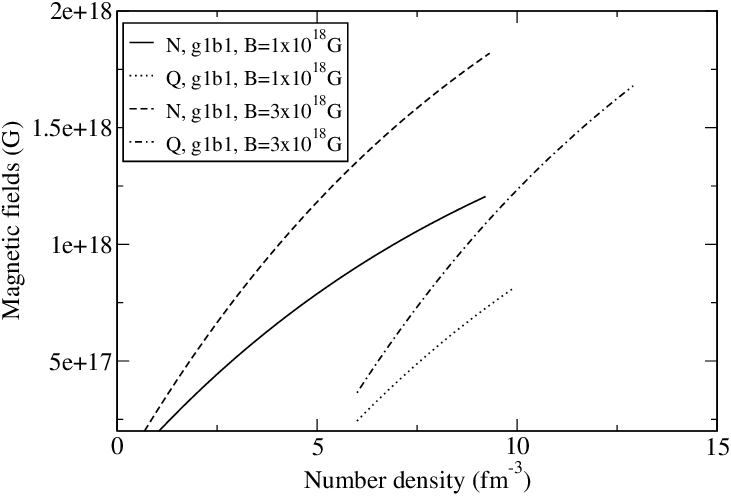}
\caption{Variation of magnetic field with number density is shown for both 
nuclear and quark matter EoS. The curves are plotted for the EoS with 
magnetic field profile corresponding to $\beta=0.1$ and $\gamma=1$.
The variation is plotted for two diffrerent ${\cal B}_C$, $1\times10^{18}$ G 
and $3\times10^{18}$ G.}
\label{mag1}
\end{figure}

\begin{figure}
\vskip 0.3in
   \centering
\includegraphics[width=2.5 in]{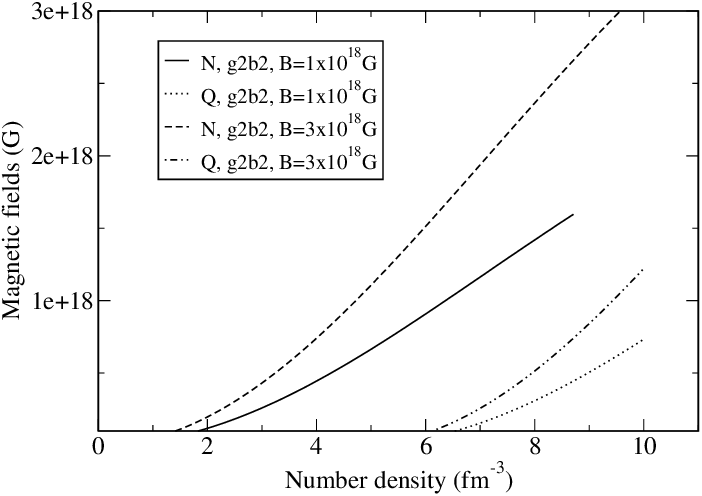}
\caption{Variation of magnetic field with number density is shown for both 
nuclear and quark matter EoS. The curves are plotted for the EoS with 
magnetic field profile corresponding to $\beta=0.2$ and $\gamma=2$.
The variation is plotted for two diffrerent ${\cal B}_C$, $1\times10^{18}$ G 
and $3\times10^{18}$ G.}
\label{mag2}
\end{figure}

From the Rankine-Hugoniot condition we can find that for the conservation 
condition to hold the number density of the SQM should be greater 
than that of the NM, but by a small amount. In fig. \ref{mag1} 
we plot 
the magnetic field variation with the number density of both matter 
phases for magnetic field profile corresponding to $\beta=0.1$ and $\gamma=1$.
For the same baryon number density the SQM can support much
lesser magnetic field than what the NM can support. Therefore, 
although there is a rise in number density due to conversion 
of NM to SQM, the magnetic field of the star reduces due to 
this conversion process. The 
conversion of NS to QS decreases the magnetic field of the 
star, resulting in a lesser magnetized QS. 
It is likely that the conversion process heats up the star, thereby 
gaining energy which is 
supplied through the conversion of magnetic energy to heat energy.
Therefore the magnetic field of the resultant QS is lesser than the 
initial NS. In fig. 
\ref{mag2} the same is 
plotted only for EoS with magnetic field
profile corresponding to $\beta=0.2$ and $\gamma=2$. The difference in the 
figures is due to same cause as discussed earlier.
 
\section{Summary} \label{summary}
We have studied the conversion of hyperon matter to SQM inside 
a NS. We have done both kinematic and dynamic study of the 
conversion process. We have seen the effect of magnetic field 
on the EoS and thereby the effect of it in the conversion process. 
We have found that beyond a certain density (on the lower side) 
the curve does not comes down. This is due to the fact that 
beyond that density the SQM is not stable. To make the conversion 
from NM to SQM possible, 
the central density of a NS must not be smaller than the minimum
density beyond which SQM is not stable. At that 
density there is a high chance of detonation wave formation. On the other 
extreme, at very high density a star may be formed but the conversion process 
would not start.  

Due to introduction of the magnetic field the range of values of baryon
density, for which the flow velocities are physical, increases. The magnetic 
field reduces the $P$ for the same
value of baryon number density, rendering the matter to be more compressible. 
That is for the same density, now the matter is less rigid or more 
compressible. The nature of the EoS depends on the magnetic field we choose, 
which depends on the central value and on the factor $\beta$ and $\gamma$. 
This $\beta$ and $\gamma$ controls the way the magnetic field would vary 
along the star for a given ${\cal B}_C$. Higher value of $\beta$ and 
$\gamma$ means the slope of the field variation is larger. Therefore 
regulating these parameters we can regulate the shock front 
propagation.

Coming to the dynamic picture, we have found that the velocity 
of the front shoots up near the center and then saturates at 
a certain velocity for higher radius. The GR effect increases 
the velocity of the front, and the rise is due to the fact 
that GR effect of the curvature of the front adds up with the 
$P$ providing a much larger thrust to the front propagation. 
As the magnetic field increases the velocity of the front decreases 
due to the negative field pressure contribution which reduces 
the effective $P$. Thus the thrust to the propagation front 
decreases, reducing the velocity. 
The most interesting fact of the study is that for such EoS if we construct 
a hyperon star and by the conversion mechanism it converts to a QS 
the magnetic field of the star decreases. Thereby signalling that the phase 
transition of NS to QS is accompanied by decrease in 
magnetic field, which goes on to heat up the star.

Although it is an interesting result but such conclusion can only be made if 
such result are consistent with direct or indirect 
observational evidences from the NS. We have performed the 
calculation for the zero temperature EoS matter, as such EoS does not have 
the provision of finite temperature inclusion. We would like to 
perform similar calculation with other  
EoS of matter having provision for inclusion of both finite temperature 
and magnetic fields. In this work the magnetic field is included through the 
EoS but we would like to perform similar 
calculation where the magnetic field is also included through the Einsteins 
field equations. Such an analysis would give a better picture of the magnetic 
effect on phase transition mechanism and which is our immediate 
agenda.

We would like to thank Dr. Banibrata Mukhopadhyay who first suggested this 
problem to us.
We would also like to thank Grant No. SR/S2HEP12/2007, funded by DST, India.
The author MS acknowledges Alexander von Humboldt Foundation for support.

\end{document}